\documentclass[preprint,showpacs,showkeywords,preprintnumbers,amsmath,amssymb]{revtex4}

\usepackage{graphicx}% Include figure files
\usepackage{dcolumn}% Align table columns on decimal point
\usepackage{bm}% bold math
\begin{document}

\title{Generalized Migdal-Kadanoff Bond-moving Renormalization Recursion Procedure I: Symmetrical Half-length Bond Operation on Translational
Invariant Lattices}

\author{Chun-Yang Wang\footnote{Corresponding author. Electronic mail:
wchy@mail.bnu.edu.cn}\footnote{Researcher in the Physical
Post-doctoral Circulation Station of Qufu Normal University}}
\author {Wen-Xian Yang}
\author {Zhi-Wei Yan}
\author {Hong Du}
\author {Xiang-Mu Kong}
\author {Yu-Qi Zhang}
\author {Ling-Yu Zhang}

\affiliation{Shandong Provincial Key Laboratory of Laser
Polarization and Information Technology, College of Physics and
Engineering, Qufu Normal University, Qufu 273165, China}

%\date{\today}

\begin{abstract}
We report in a series of papers two types of generalized
Migdal-Kadanoff bond-moving renormalization group transformation
recursion procedures. In this first part the symmetrical operation
of half length bonds on translational invariant lattices are
considered. As an illustration of their predominance in application,
the procedures are used to study the critical behavior of the
spin-continuous Gaussian model constructed on the triangular
lattices. Results such as the correlation length critical exponents
obtained by this means are found to be in good conformity with the
classical results from other studies.
\end{abstract}

%\keywords{bond-moving renormalization; Critical exponent; Gaussian model}

\pacs{05.50.+q,64.60.-i,75.10.Hk}

\maketitle

\section{INTRODUCTION}

In the early 1970s a very important contribution to the calculation
of near-critical properties in many-particle systems was made by A.
A. Migdal in which a set of renormalization group recursion formulas
were proposed basing on a series of bond-moving operations
\cite{Migdal1,Migdal2}. Soon after its publication, L. P. Kadanoff
consummated Migdal's remarkable work to include potential-moving
operations \cite{Kadanoff}. This enables the bond-moving procedure
together with the decimation \cite{Kadanoff2,Nelson} and block
transformation
\cite{Kadanoff3,Neimeijer,Nauenberg,Kadanoff4,Kadanoff5,Bell} to
form three pillars of coarsening methods in the study of phase
transition and critical phenomena via renormalization group theory.

The bond-moving procedure is important because it can be applied to
systems with almost all kinds of internal symmetries. The results
got from them are asymptotically exact and can be analytically
continued in dimensionality with relative ease in the limits of both
weak and strong couplings \cite{Wilson,Balkan}. With these
advantages, it has offered the possibility of attacking
strong-coupling phase transitions in giving an exact determination
of the critical point of the two-dimensional Ising system with dual
symmetries \cite{Kramers,Wegner} and has been an effective starting
point for dealing with highly anisotropic systems such as the
Heisenberg model at dimensionality $2+\epsilon$.

Effective as it has always been, the Migdal-Kadanoff bond-moving
recursion procedure is recently found not as consummate as it should
have been. For example, it is preferred in dealing with
spin-discrete systems constituted on traditional globally symmetric
lattices. For systems with local symmetries such as some fractals it
appears to be sometimes inconvenient or even powerless. For these
reasons, we aim in this paper to improve the Migdal-Kadanoff
bond-moving recursion procedure to seek a more expanded rang of its
application.

The paper is organized as follows: in Sec. \ref{sec2}, we give the
improvement we have made on the procedures by recurring it on the
triangular lattices; in Sec. \ref{sec3}, the critical behavior of
the spin-continuous Gaussian model is studied by using of the newly
generalized procedures as an illustration of their predominance in
application. Sec. \ref{sec4} serves as a summary of our conclusion
in which further implicit applications are also discussed.

\section{generalization}\label{sec2}

In this section, we will report the improvement we made on the
bond-moving procedure by recurring them on the triangular lattice
which is a simple but typical configuration with globally symmetry
and the spin systems constructed on it can be easily treated with
various renormalization methods besides bond-moving.

\begin{figure}
%\centering
\includegraphics[scale=0.85]{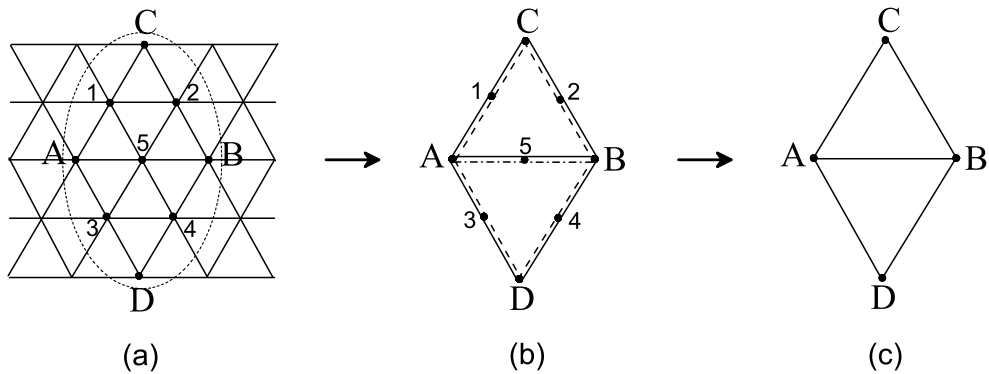}
\caption{Traditional bond-moving procedure used on the triangular
lattice in previous studies where the sites $i=1,2,\cdots,5$ in (a)
and (b) are those to be decimated. \label{tra-pro}}
\end{figure}

For comparision, we firstly present the traditional bond-moving
procedure generally adopted on the triangular lattice. It proceeds
as follows: (1) selecting a cluster of lattice sites as is shown in
fig.\ref{tra-pro}a; (2) moving the bonds between each pair of to be
eliminated sites $i=1,2,\cdots,5$ to the peripheral position (Seen
in fig.\ref{tra-pro}b for an example) to make the lattice
coarse-grained; (3) rescaling the system and decimating the to be
eliminated sites. Thus it results in a good approximation for the
original lattice system and opens up a convenient way for studying
the critical properties of the spin systems constructed on it.

However, although it is effective and gives reasonable
results for many spin-discrete systems, the traditional bond-moving
procedure feels powerless in dealing with the spin-continuous
systems such as Gaussian and $S^{4}$ models. This is because the
cluster selected in step (1) is relatively too large (including five
sites to be eliminated) and this results in several complicated
integrations (at least five fold of integration) to be treated in
the consecutive site-decimating procedure in step (3). Furthermore,
in order to keep the globally symmetry unchanged, the moving of the
bond between sites 1 and 2 (as well as 3 and 4) must be operated in
a way different from other bonds. This reveals some irregularity of
the bond-moving recursion procedure.

\begin{figure}
%\centering
\includegraphics[scale=0.8]{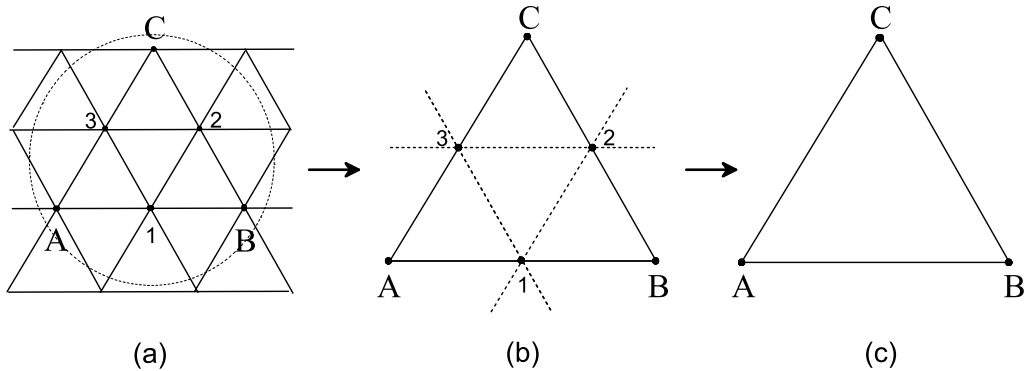}
\caption{Improved bond-moving procedure recurring on the triangular
lattice where in (a) and (b) the peripheral bonds connecting the
three to be eliminated sites 1, 2 and 3 around the selected triplet
$\triangle ABC$ are drawn at half length. \label{pro1}}
\end{figure}

Basing on these considerations, we present here an improved
 bond-moving recursion method to avoid the occurrence
of these incongruities. It is proceeded in such a little different
way: (1) selecting a cluster of only six lattice sites as a basic
unit for recursion where the to be moved peripheral bonds connecting
the three to be eliminated sites 1, 2 and 3 around the selected
triplet $\triangle ABC$ are regulated to be in the same order of
total length by including two half length bonds as is shown in
fig.\ref{pro1}b; (2) moving the bonds connecting each pair of to be
eliminated sites $i=1,2,3$ to the peripheral position to make the
lattice coarse-grained; (3) rescaling the system and decimating the
to be eliminated sites. This recursion procedure can be proved to be
a more powerful way bringing with great convenience in the study of
spin-continuous systems.

\section{predominance illustration} \label{sec3}

In order to make a display of its predominance in application, we
present here an illustrative study on the critical behavior of the
spin-continuous Gaussian model constructed on the triangular-lattice
by using of our improved bond-moving recursion.

Let us begin with the classical Gaussian effective Hamiltonian
\begin {equation}
H_{\textrm{eff}}=\sum_{\langle
ij\rangle}K\sigma_{i}\sigma_{j}-\frac{b}{2}\sum_{i}\sigma_{i}^{2},
\end {equation}
where $\langle ij\rangle$ in the summation represents a certain
nearest-neighbor spin pair; $K={J}/{k_{B}T}$ is the reduced
interaction with $K>0$ denotes the ferromagnetic system; $b$ is the
Gaussian distribution constant; $k_{B}$ the Boltzmann constant and
$T$ the thermodynamic temperature. The spins can take any real value
between $(-\infty,+\infty)$ and the probability of finding a given
spin between $\sigma_{i}$ and $\sigma_{i} + d \sigma_{i}$ is assumed
to be the Gaussian-type distribution
$p(\sigma_{i})d\sigma_{i}\propto[\exp-(b/2)\sigma_{i}^{2}]
d\sigma_{i}$.

\begin{figure}
\centering
\includegraphics[scale=1.2]{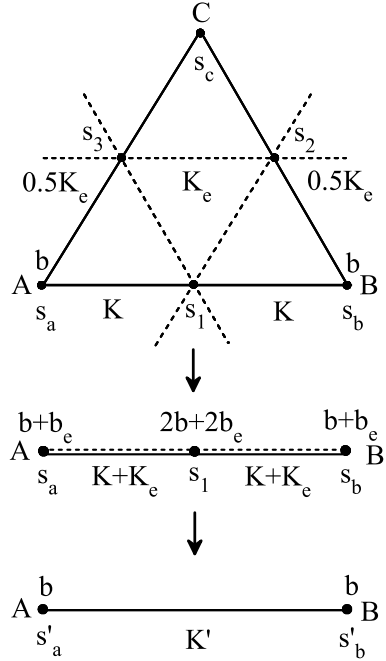}
\caption{Bond-moving and decimation procedure of
renormalization-group transformation for the renormalized bond $K'$
between sites A and B of the Gaussian model constructed on the
triangular lattice.\label{pro2}}
\end{figure}

Fig.\ref{pro2} gives the bond-moving and decimation approach to the
Gaussian model on the triangular-lattice performed as Gefen et al
did on the fractal \cite{Genfen}. Where the peripheral bonds
connecting sites 2 and 3 around the selected triplet are drawn at
half length and two types of interactions $K_{e}$ and $K$ together
with two types of self-energy $(-b_{e}s^{2}/2)$ and $(-bs^{2}/2)$
are assigned respectively for differentiation of the to be and not
to be eliminated bonds. For the particular case of triangular
lattice the numerical value of $K_{e}$ and $K$ is actually identical
as well as that of $b_{e}$ and $b$. In the bond-moving procedure the
two half length bonds are considered acting effectively as a whole
one and be moved regularly as other bonds. The decimation procedure
for the renormalized bond $K'$ is
\begin{eqnarray}
&&\int^{+\infty}_{-\infty}\textrm{exp}\left[\left(K+K_{e}\right)\left(s_{a}s_{1}+s_{1}s_{b}\right)
-\frac{b+b_{e}}{2}\left(s^{2}_{a}+s^{2}_{b}\right)-\frac{2b+2b_{e}}{2}s^{2}_{1}
\right]ds_{1}
\nonumber\\&&\hspace{1.0cm}=\int^{+\infty}_{-\infty}\textrm{exp}\left[2K\left(s_{a}s_{1}+s_{1}s_{b}\right)
-b\left(s^{2}_{a}+s^{2}_{b}\right)-2bs^{2}_{1}\right]ds_{1}
\nonumber\\&&\hspace{1.0cm}=C\textrm{exp}\left[K's'_{a}s'_{b}-\frac{b}{2}\left(s'^{2}_{a}+s'^{2}_{b}\right)\right]
\label{decimation},
\end{eqnarray}
where the relations of $K=K_{e}$ as well as $b=b_{e}$ are used.

By directly integrating $s_{1}$ to decimate the intermediate spins
it becomes
\begin{eqnarray}
C\textrm{exp}\left[\frac{K^{2}}{b}s_{a}s_{b}-\left(b-\frac{K^{2}}{2b}\right)\left(s^{2}_{a}+s^{2}_{b}\right)\right]
=C\textrm{exp}\left[K's'_{a}s'_{b}-\frac{b}{2}\left(s'^{2}_{a}+s'^{2}_{b}\right)\right]
\label{integrate},
\end{eqnarray}
For the continuity of spin sampling, we can rescale the renormalized
spins by
\begin{eqnarray}
s'_{a}=\xi_{a}s_{a} \hspace{0.8cm} \textrm{and} \hspace{0.8cm}
s'_{b}=\xi_{b}s_{b} \label{rescale}
\end{eqnarray}
with
\begin{eqnarray}
\xi^{2}_{a}=\xi^{2}_{b}=2-\frac{K^{2}}{b^{2}}\label{rescalefactor}.
\end{eqnarray}
Then the recursion relation for $K'$ reads
\begin{eqnarray}
K'=R(K,K_{e})=\frac{1}{\xi_{a}\xi_{b}}\frac{K^{2}}{b}=\frac{K^{2}b}{2b^{2}-K^{2}}\label{rescalefactor}.
\end{eqnarray}
The critical point is obtained to be $K=b$ and the
renormalization-group transformation matrix at this point reduced to
$1\times1$ with only one eigenvalue
\begin{eqnarray}
\lambda=\frac{\partial K'}{\partial
K}|_{K=b}=\frac{4b^{3}K}{(2b^{2}-K^{2})^{2}}|_{K=b}=4\label{eigenvalue}.
\end{eqnarray}
Thus we obtain the critical exponent of correlation length
\begin{eqnarray}
\nu=\frac{\textrm{ln}B}{\textrm{ln}\lambda}=\frac{\textrm{ln}2}{\textrm{ln}4}=0.5\label{criticalE}
\end{eqnarray}
where the triangular lattice is treated as a special kind of fractal
with subdividing factor $B=2$. Here we can found that the results we
have got is in good conformity with the previous ones
\cite{chyw1,chyw2,zqlin}. However the recursion procedure is quite
simple with only one fold of integration need to be performed.

\begin{figure}
\centering
\includegraphics[scale=1.2]{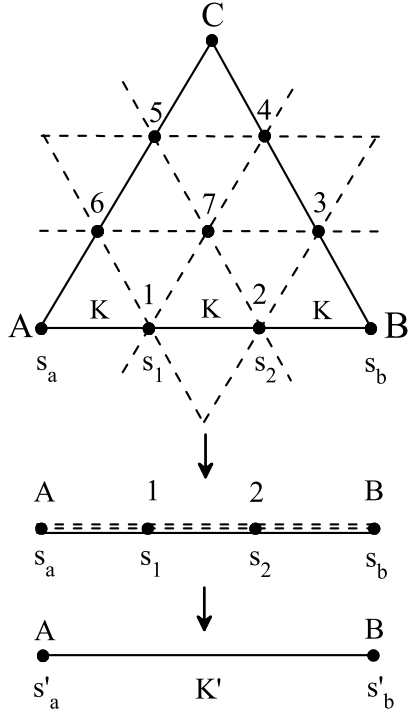}
\caption{Bond-moving and decimation procedure for the renormalized
bond $K'$ between sites A and B of the Gaussian model constructed on
the triangular lattice in the case of considering the next
nearest-neighbor interactions.\label{pro3}}
\end{figure}

This is not the only case to illustrate its predominance in
application. In fact, more convenience can be found in treating some
more complicated cases. For example, Fig \ref{pro3} presents the
bond-moving and decimation procedure in the case of considering next
nearest-neighbor interactions. From which we can see that although
the number of to be eliminated sites increases to six, the
integration that needs to be performed in the decimation increases
only to two folds. This brings little numerical difficulty for the
study of spin-continuous models.

Table \ref{table1} reports the related results such as the critical
point and correlation length critical exponent of the Gaussian model
we obtained in the case of considering two types of long-rang
decaying next nearest-neighbor interactions respectively. Where the
power law decaying is assumed to be
\begin{eqnarray}
K(r_{mn})=K\left(\frac{r_{mn}}{a}\right)^{-\alpha}\label{power},
\end{eqnarray}
while the exponential decaying is supposed to be
\begin{eqnarray}
K(r_{mn})=Ke^{-\left(\frac{r_{mn}}{a}-1\right)\alpha}\label{power},
\end{eqnarray}
in which $a$ is the lattice constant, $\alpha$ the decaying
exponent; $K(r_{mn})$ represents the reduced interaction between
spin pairs $m$ and $n$ separated by a distance $r_{mn}$; when
$r_{mn}=a$, $K(r_{mn})$ reduced to the nearest neighbor interaction
$K$.

\begin{table}
\caption{\label{table1} The related results of the Gaussian model
constructed on the triangular lattice in the case of considering
next nearest-neighbor interactions. Where $K_{c}$ denotes the
critical point, $\lambda$ is the eigenvalue of the transformation
matrix, $\nu$ the correlation length critical exponents of the
Gaussian model.}
\begin{ruledtabular}
\begin{tabular}{lcccccc}
 $\alpha$ &$K_{c}$\footnotemark[1] &$\lambda$\footnotemark[1] &$\nu$\footnotemark[1]
 &$K_{c}$\footnotemark[2] &$\lambda$\footnotemark[2] &$\nu$\footnotemark[2]\\
\hline
                  1 & 0.8828b & 6.499 & 0.5870 & 0.9123b & 6.989 & 0.5650 \\
                  2 & 0.9395b & 7.511 & 0.5449 & 0.9667b & 8.117 & 0.5246 \\
                  3 & 0.9692b & 8.178 & 0.5228 & 0.9876b & 8.651 & 0.5092 \\
                  4 & 0.9845b & 8.566 & 0.5114 & 0.9954b & 8.868 & 0.5034 \\
                  5 & 0.9922b & 8.775 & 0.5058 & 0.9983b & 8.949 & 0.5013 \\
                  6 & 0.9961b & 8.887 & 0.5029 & 0.9994b & 8.984 & 0.5004 \\
$\rightarrow\infty$ & b       & 9.000 & 0.5000 & b       & 9.000 & 0.5000 \\
\end{tabular}
\end{ruledtabular}
\footnotetext[1]{power law decaying} \footnotetext[2]{exponential
decaying}
\end{table}%7 & 0.9980b & 8.937 & 0.5016 & 0.9998b & 8.997 & 0.5001 \\

Seen from Table \ref{table1} again that the results we have got in
the case of considering next nearest-neighbor interactions are also
in good conformity with the previous studies. But the bond-moving
and decimation recursion procedure we have used is quite simple than
the usual means reviewed in Sec \ref{sec2}. This reveals the
improved bond-moving renormalization group transformation procedure
we reported in this paper is trustworthy and can bring great
convenience in future applications.

\section{summary and discussion} \label{sec4}

In summary, we have presented in this paper an improved
Migdal-Kadanoff bond-moving recursion renormalization group
transformation procedure including half length operations. Results
obtained by this means such as the correlation length critical
exponents of the traditional Gaussian model constructed on the
triangular lattices are found to be in good conformity with the
classical results from other studies. In particular, the included
half length operation is revealed to be able to bring with great
conveniences for the study of complicated spin-continuous systems
which illustrates a remarkable predominance in application of our
generalized Migdal-Kadanoff bond-moving recursions. With these
advantages, the future applying of this method on more complicated
spin systems such as the $S^{4}$ model and even some fractal systems
is worth to be eagerly expected.

\section * {ACKNOWLEDGEMENTS}

This work was supported by the Shandong Province Science Foundation
for Youths (Grant No.ZR2011AQ016), the  Shandong Province
Postdoctoral Innovation Program Foundation (Grant No.201002015), the
Scientific Research Starting Foundation, Youth Foundation (Grant
No.XJ201009) and the Foundation of Scientific Research Training Plan
for Undergraduate Students (Grant No.2010A023) of Qufu Normal
University.

\end{document}